# Thermal and Mechanical Stability of Zeolitic Imidazolate Frameworks Polymorphs

Lila Bouëssel du Bourg,[1] Aurélie U. Ortiz,[1] Anne Boutin,[2] and François-Xavier Coudert[1, a)]

[1]*PSL Research University, Chimie ParisTech – CNRS, Institut de Recherche de Chimie Paris, 75005 Paris, France*

[2]*École Normale Supérieure, PSL Research University, Département de Chimie, Sorbonne Universités – UPMC Univ Paris 06, CNRS UMR 8640 PASTEUR, 24 rue Lhomond, 75005 Paris, France*

(Dated: December 19, 2014)

Theoretical studies on the experimental feasibility of hypothetical Zeolitic Imidazolate Frameworks (ZIFs) have focused so far on relative energy of various polymorphs, by energy minimization at the quantum chemical level. We present here a systematic study of stability of 18 ZIFs as a function of temperature and pressure, by molecular dynamics simulations. This approach allows us to better understand the limited stability of some experimental structures upon solvent or guest removal. We also find that many of the hypothetical ZIFs proposed in the literature are not stable at room temperature. Mechanical and thermal stability criteria thus need to be considered for the prediction of new MOF structures. Finally, we predict a variety of thermal expansion behavior for ZIFs as a function of framework topology, with some materials showing large negative volume thermal expansion.

## I. INTRODUCTION

Zeolitic imidazolate frameworks (ZIFs) are a subclass of metal–organic frameworks (MOF) that feature imidazolate linkers bridging metal centers to form three-dimensional porous crystalline solids isomorphous to zeolitic frameworks.[1–4] ZIFs have recently gained considerable attention for their potential applications, e.g. in domains such as $CO_2$ capture,[5] sensing,[6] encapsulation and controlled delivery,[7] and fluid separation.[8–11] ZIFs as a family are often thought of as having specific advantages over MOFs in general. It is often stated that they inherit desirable qualities from both the MOF and zeolite worlds: the tunable porosity, structural flexibility and the functionalization of the internal surface of the MOFs, as well as the thermal, mechanical and chemical stability of the zeolites.

Moreover, the topological equivalence between the metal–imidazolate four-fold coordination chemistry and the corner-sharing $SiO_4$ tetrahedra from which zeolites are built means that many ZIF topologies can potentially be synthesized. 218 zeolitic topologies are known to date,[12] out of an infinity of mathematically possible periodic four-connected nets. And indeed, in still very recent field of ZIFs, over a hundred of different ZIF structures have been reported so far in the literature, either by direct solvothermal synthesis, mechanochemistry,[13] solvent-assisted linker exchange,[14] or transmetalation.[15] In addition, experimental investigations of the relative stabilities of ZIFs,[16] as well as theoretical calculations on hypothetical ZIF structures, have shown that many ZIF polymorphs fall within a small energy range of experimentally synthesized structures.[17,18] This has naturally lead to the conclusion than many of these ''undiscovered nanoporous topologies should be amenable to synthesis''.[17]

However, while the overall number of ZIFs and ZIF-like structures experimentally realized continues to increase, the number of topologies accessible *for a given linker such as unsubstituted imidazolate* seem to be rather limited. In particular, it appears

---

a)Electronic mail: fx.coudert@chimie-paristech.fr

that number is much smaller than the number of hypothetical structures which would be considered *experimentally feasible*, based on considerations on energy alone. Indeed, energetic considerations neglect other key components in terms of stability in real conditions: the effect of solvent (e.g. during the solvothermal synthesis), thermal motion and entropy (forcing one to compare not only relative energies, but relative free energies), and behavior under external mechanical constraints (such as isotropic pressure and shear stresses, both of which can occur in practical applications). Such factors have been little studied in the literature so far, with the notable exception of the recent work by Gee and Sholl[19] studying the influence of solvent and temperature, although the later was only treated in the harmonic approximation. This is a severe limitation, especially in framework materials such as ZIF which display many low-frequency vibrations modes, with strong anharmonicity.

We report in this paper the first systematic investigation of the thermal and mechanical stability of porous ZIFs, in order to bring further insight into the question of feasibility of this topical family of metal–organic frameworks. Our earlier work on the mechanical stability of ZIF-8,[20] as well as on the *ab initio* prediction of structural transitions in flexible metal–organic frameworks,[21,22] has shown how molecular simulation can shed light into the stability (or lack thereof) of MOFs as a function of both temperature and pressure. We now extend this proven methodology to the systematic study of an entire family of materials, both experimentally observed and hypothetical, in order to provide a deeper understanding of the thermal and mechanical components of ZIF stability.

## II. METHODS

Molecular dynamics (MD) simulations of various ZIF frameworks, empty and with $CH_4$ molecules loaded in their pores, were performed in the isostress-isothermal ensemble $(N, \sigma, T)$ using the NAMD 2.9 software package.[23] The temperature was fixed using Langevin dynamics on the heavy atoms, with a damping coefficient of 10 $ps^{-1}$. The pressure was fixed by a modified Nosé-Hoover method, which is a combination of the constant pressure algorithm proposed by Martyna et al.[24] with piston fluctuation control implemented using Langevin dynamics.[25] We used a piston oscillation period of 0.2 ps and a piston decay time of 0.1 ps.

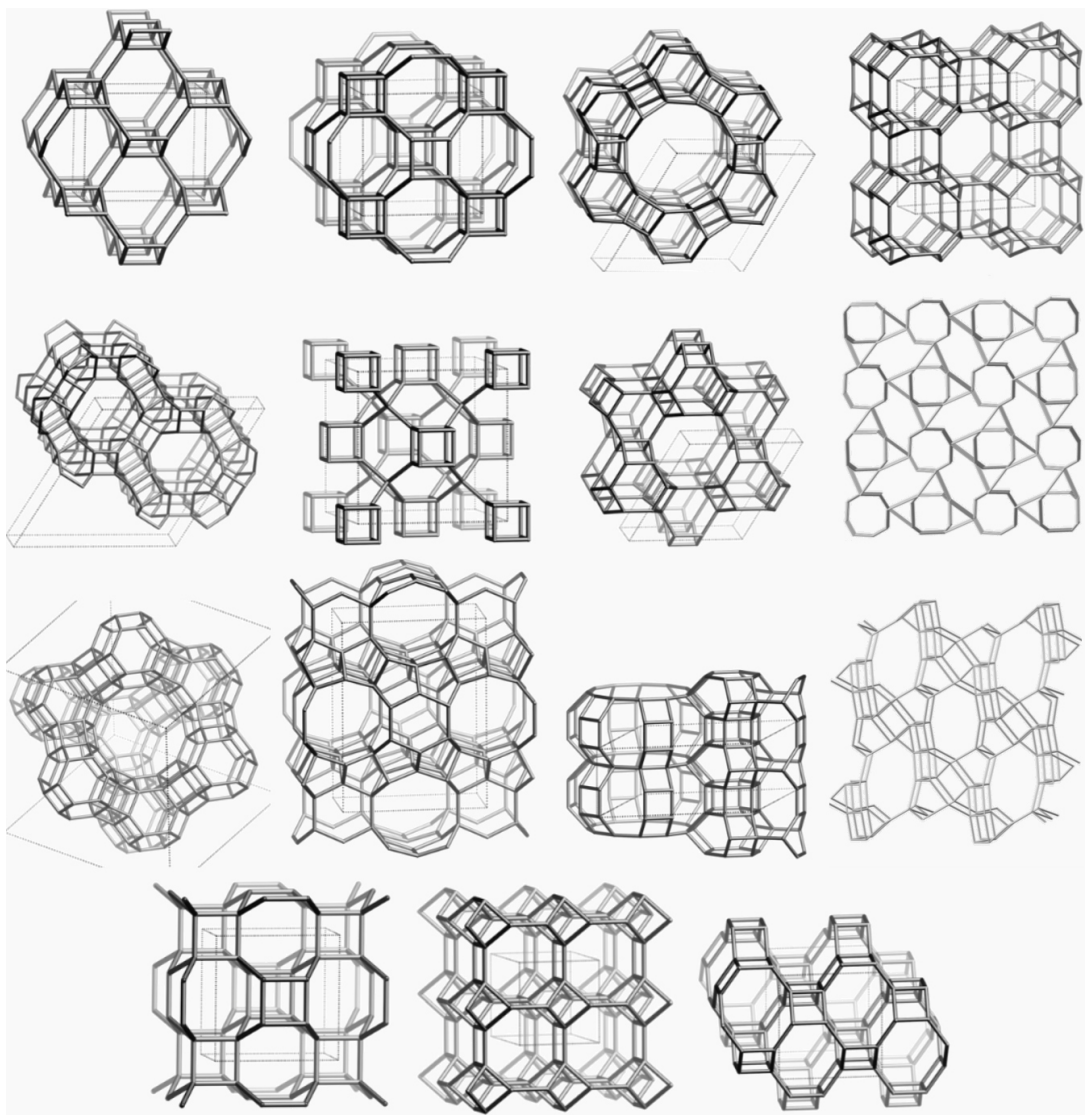

Figure 1. Framework topologies studied in this work. From left to right and top to bottom: **ABW**, **ACO**, **AFI**, **ATN**, **ATO**, **AST**, **CAN**, **coi**, **FAU**, **FER**, **LTL**, **nog**, **BCT**, **DFT** and **GIS**.

Though the stress considered in this work corresponded to isotropic compression, the simulation was performed in the $(N, \sigma, T)$ ensemble, and not the $(N, P, T)$ ensemble, by allowing full flexibility of the unit cell. The NAMD source code for the barostat was patched in order to allow the unit cell to be fully flexible, with random variations of all components of the unit cell vectors (rather than the 3 vector lengths as implemented in NAMD version 2.9). The patch is available as supplementary material.[26]

An integration time step of 1.0 fs was used, and each MD simulation was run for 5 ns, of which the first 1 ns was discarded as an equilibration period and not used for the calculation of averages and time correlations. Checks performed with longer simulations showed that calculation of elastic constants from 4 ns of unit cell fluctuations in the $(N, \sigma, T)$ ensemble yielded elastic constants with an uncertainty of ±0.03 GPa. In all cases, we checked that the elastic tensor is definite positive, and thus fulfills the Born elastic stability criterion.[27]

All simulations were run on a supercells of the respective ZIF structures, so that each dimension of the simulation box was between 40 and 60 Å. Full three-dimensional periodic boundary conditions were employed. Electrostatic interactions were treated using the particle mesh Ewald (PME) method, and a cut-off distance of 14 Å was used for the summation of Lennard-Jones interactions.

The force field used to describe intra- and intermolecular interactions of the ZIF structures, as well as the ZIF/CH$_4$ inter-molecular interactions, were taken from the very recent work of Zhang et al.[28] This force field is derived from the very generic AMBER forcefield, where some of the terms involving Zn atoms were reoptimized to better reproduce the ZIF-8 structure and possible overall rotations of the imidazolate linker. Since ZIF-8 contains 2-methylimidazolate as a linker, we adapted it slightly by replacing the methyl group with a hydrogen atom, taking its parameters from the original AMBER parametrization for consistency. The reasonable agreement found with experimental structures (geometry and lattice parameters) for the stable ZIFs, as described later in the text, validates this generic approach.

| Topology | Zeolitic | Exptal | Name | Ref. |
|---|---|---|---|---|
| AFI | yes | no | | 17 |
| CAN | yes | no | | 17 |
| cag | no | yes | ZIF-4 | 1 |
| coi | no | yes | | 31 |
| DFT | yes | yes | ZIF-3 | 1 |
| FAU | yes | no | | 17 |
| LTL | yes | no | | 17 |
| MER | yes | yes | ZIF-10 | 1 |
| nog | no | yes | | 32 |
| SOD | yes | yes | SALEM-2 | 14 |
| ATN | yes | no | | 17 |
| ATO | yes | no | | 17 |
| FER | yes | no | | 17 |
| ABW | yes | no | | 17 |
| ACO | yes | no | | 17 |
| AST | yes | no | | 17 |
| BCT | yes | yes | ZIF-1 | 1 |
| GIS | yes | yes | ZIF-6 | 1 |

Table I. List of all Zeolitic Imidazolate Framework Zn(imidazolate)$_2$ polymorphs studied in this work. For zeolitic topologies, the code indicated is the International Zeolite Association[12] three-letter uppercase code; for nonzeolitic topologies, the RCSR code.[30]

## III. RESULTS AND DISCUSSION

In order to carry out a systematic study of Zeolitic Imidazolate Frameworks' behavior as a function of temperature and pressure, we performed molecular dynamics simulations of 18 different ZIFs, all polymorphs of Zn(*im*)$_2$ composition (where *im* is unsubstituted imidazolate) with various topologies. The full list of systems studied is given in Table I. It includes 8 ZIF frameworks experimentally synthesized in this composition and reported in the literature.[29] Five of those have a zeolitic topology, i.e. correspond to one of the 218 four-connected tetrahedral topologies enumerated by the International Zeolite Association (**BCT**, **DFT**, **GIS**, **MER**, and **SOD**).[12] The other three also feature a four-connected net, but one that does not correspond to any of the zeolitic topology. These are designated by their Reticular Chemistry Structure Resource (RCSR) net code[30], in lower case: **cag**, **coi**, and **nog**.

We have also included in our study 10 hypothetical ZIF frameworks, which have been artificially created from the corresponding zeolitic structures (**ABW**, **ACO**, **AFI**, **AST**, **ATN**, **ATO**, **CAN**, **FAU**, **FER**, and **LTL**). Lewis et al., who studied the energetic stability of these hypothetical frameworks through quantum chemical calculations, found that all ten were metastable polymorphs of Zn(*im*)$_2$ at zero temperature, i.e. local minima in energy.[17] Moreover, many of those are within a small energy range from the most stable polymorph, the nonporous ZIF of topology **zni**.[33] It is then natural to assume that many of these polymorphs should be experimentally feasible on the basis of their low enthalpy of formation.[16–18] They have not, however, been experimentally synthesized *with an unsubstituted imidazolate linker* since this prediction six years ago.

### A. Stability at ambient conditions

We first performed molecular dynamics simulations of the 18 above-listed polymorphs in the absence of external mechanical pressure ($\sigma = 0$), at temperatures between 77 and 400 K. All simulations were started from the "ideal" structures, which are either the crystallographic structures, or the quantum chemical energy minimized structures (for experimental and hypothetical frameworks respectively). From the evolution of unit cell parameters (see Figure 2) and unit cell volume, as well as the visualization of the geometry of the ZIF framework itself, three categories of behavior were observed:

(i) frameworks that are stable in the whole temperature range (up to 400 K): **AFI**, **coi**, **FAU**, **nog**, **MER**, **DFT**, **cag**, **SOD**, **LTL**, and **CAN**;

(ii) frameworks stable only at low temperature ($T < 300$ K): **ATN**, **FER** and **ATO**.

(iii) frameworks that are not stable at any temperature in the 77–400 K range, i.e. that spontaneously undergo a transition into a different structure during our MD simulations.

As seen in Fig. 2, the materials of the first group show very little influence of temperature. The materials in this group include some experimentally known materials, including the widely studied **SOD** and **cag** ZIFs (SALEM-2 and ZIF-4, respectively), as well as some hypothetical structures (**AFI**, **CAN**, **FAU** and **LTL**). Their structural features (checked by visual inspection) are identical to the structures calculated by quantum chemistry, and the unit cell parameters only differ to a small extent (up to 10%). This "deformation" is indicative of the difference in description between the quantum-chemical description and the structure given using the force field approximation.[34]

The second group is composed of hypothetical frameworks (**ATN**, **ATO** and **FER**) that have limited thermal stability. These predicted structures, which are of relatively low energy, are indeed stable at low temperature (77 K), but not at room temperature: they undergo a spontaneous transition into another phase, in relatively short timescales (less than a nanosecond). This provides a good explanation for the fact that, although deemed "experimentally feasible" on the basis of their formation enthalpy, no synthesis of these three materials has been reported in the literature. It thus underlines one of the drastic limitations of using zero-Kelvin energy calculations (typically at the Density Functional Theory level) for the computational screening of new materials: although this approach is really good at rationalizing the experimental formation of different topologies as a function of metal and linker modifications,[35,36] its predictive power is limited because it addresses only one of the key criteria for stability of a material.

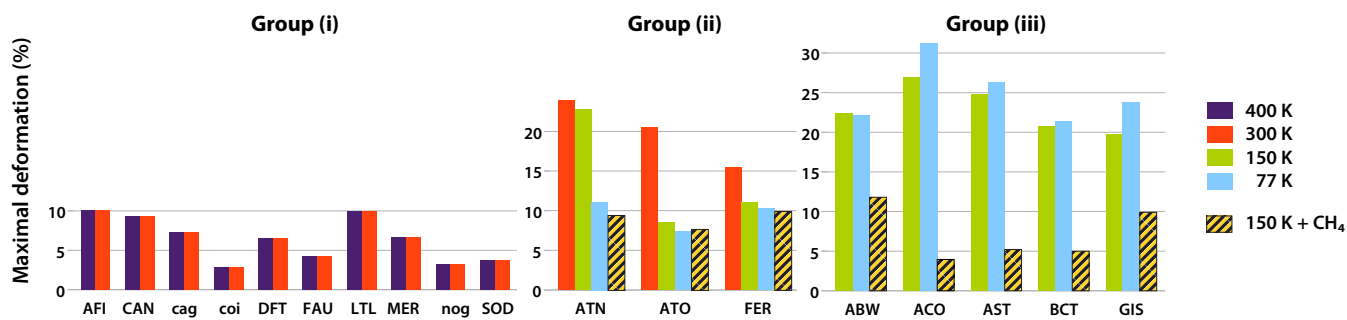

Figure 2. Maximal deformation of the ZIF structures studied (in percentage of unit cell parameters), in the 77–400 K temperature range, taking as a reference the structures predicted by quantum chemistry calculations.[17] The frameworks are classified in three categories (see text for details). Panels *(ii)* and *(iii)* also feature results in the presence of methane adsorbed inside the pores, at 150 K.

The structures of the third group exhibit a more drastic behavior: they are only stable in the limit of zero temperature ($T \to 0$), but mechanically unstable at temperature even as low as 77 K. In order to check that their lack of stability was not due to a bad description of these structures by the force field used, we performed energy minimizations and checked that the resulting configurations were close to the "ideal" structures. So, these structures represent local minima in the potential energy surface of $Zn(im)_2$ (both at the DFT level and with the force field description). Nevertheless, these minima are so shallow that thermal motions at even low temperature (77 K) allow the system to escape them. Thus, they do not correspond to metastable states of the framework at finite temperature, i.e. they are not local minima of the free energy.

What seems puzzling, however, is that this list contains both hypothetical structure (**ABW**, **ACO**, and **AST**) but also experimentally reported ZIF structures (**BCT** = ZIF-1, **GIS** = ZIF-6). However, these structures have been initially reported "as synthesized"[1], with solvent molecules (typically dimethylformamide or diethylformamide) still inside the pores. We haven't been able to find in the literature a single example of ZIF-1 or ZIF-6 being evacuated or activated, nor used for adsorption, catalysis or any other application. We thus predict on the basis of our simulations that, like some other reported ZIFs,[14] ZIF-1 and ZIF-6 are not stable upon removal of guest molecules. We furthermore predict that the same is true of the hypothetical **ABW**, **ACO** and **AST** frameworks.

In order to confirm this effect of guest molecules directly through molecular simulation, we performed an additional series of MD simulations with guest molecules adsorbes inside the pores of the frameworks. We used methane as a generic guest for these simulations, as it has no strong or specific interactions with the ZIF framework and will just show the effect of pore filling on the mechanical stability. For each framework, we performed Grand Canonical Monte Carlo (GCMC) calculations of methane adsorption to find the methane saturation uptake, and then ran molecular dynamics simulations in the isostress-isothermal ensemble ($N, \sigma, T$), i.e. with fixed quantity of methane adsorbed, at 150 K and in the absence of external pressure. The results, showed in Fig. 2(iii), clearly demonstrate that all five frameworks are now mechanical stable in the presence of guests, in agreement with the experimental observations. It also shows that this stabilization is dominated by a strong and generic pore filling effect, even in the absence of specific host–guest interactions (as is the case here with methane), in line with earlier work on the topic.[20,37] The same is true of frameworks of group (ii), especially **ATN** which was not stable without guest molecules at 150 K. For frameworks of group (i), however, there was no major effect: only the standard adsorption-induced variation of volume (of the order of 0.2%, and thus not depicted on Fig. 2) was observed.

**B. Stability under pressure**

We then turned our attention to the stability of ZIF frameworks under pressure. In the past, a lot of work has focused on the mechanical properties (elastic moduli, hardness) of MOFs and ZIFs, both experimentally[38–40] and computationally.[41,42] The occurence of pressure-induced crystal-to-crystal and crystal-to-amorphous transitions has been solidly established in several frameworks, including ZIF-8,[43] ZIF-4[44] and the non-porous ZIF-zni.[45] Here, we try to shed light into the generality of this pressure-induced transitions by our systematic approach, modelling the behavior under pressure of 10 different ZIFs with identical chemical composition ($Zn(im)_2$), at the same level of molecular modelling.

Starting from the 10 structures stable at room temperature (group *(i)* in Fig. 2), we performed series of MD simulations at increasing values of pressure, mimicking an hydrostatic compression of the material by a nonpenetrating fluid. From these series of simulations at 0.1 GPa intervals, as well as an additional point at 0.05 GPa, we obtain the limits of stability for each of the frameworks studied, reported in Table II. First, we conclude that most ZIFs studied show relatively low stability upon compression, compared to inorganic materials or molecular framework materials, who usually resist to GPa-scale pressures. The exception here is the **coi** framework, which is very dense (porosity of 9% for a probe of radius 1.2 Å). All other frameworks are much less stable, including some like the very porous **FAU** that are stable at ambient conditions, but unstable even the very modest pressure of 50 MPa. This is in line with the little experimental data available, namely the pressure-induced amorphization of ZIF-8 at 3.4 GPa,[43] and the structural transition of ZIF-4 in the range of 0.12–0.21 GPa (depending on the pressure-transmitting fluid used for the compression).[44] Apart from the special case of **coi** (very dense and very stable), there is no clear link between the framework's density (or porosity) and its stability under pressure, unlike the correlation Tan et al. observed between ZIFs' elastic moduli and porosity,[38] and the similar behavior known in zeolites.[46] It appears that mechanical stability of ZIFs depends in a more intricate way on the details of the framework topology.

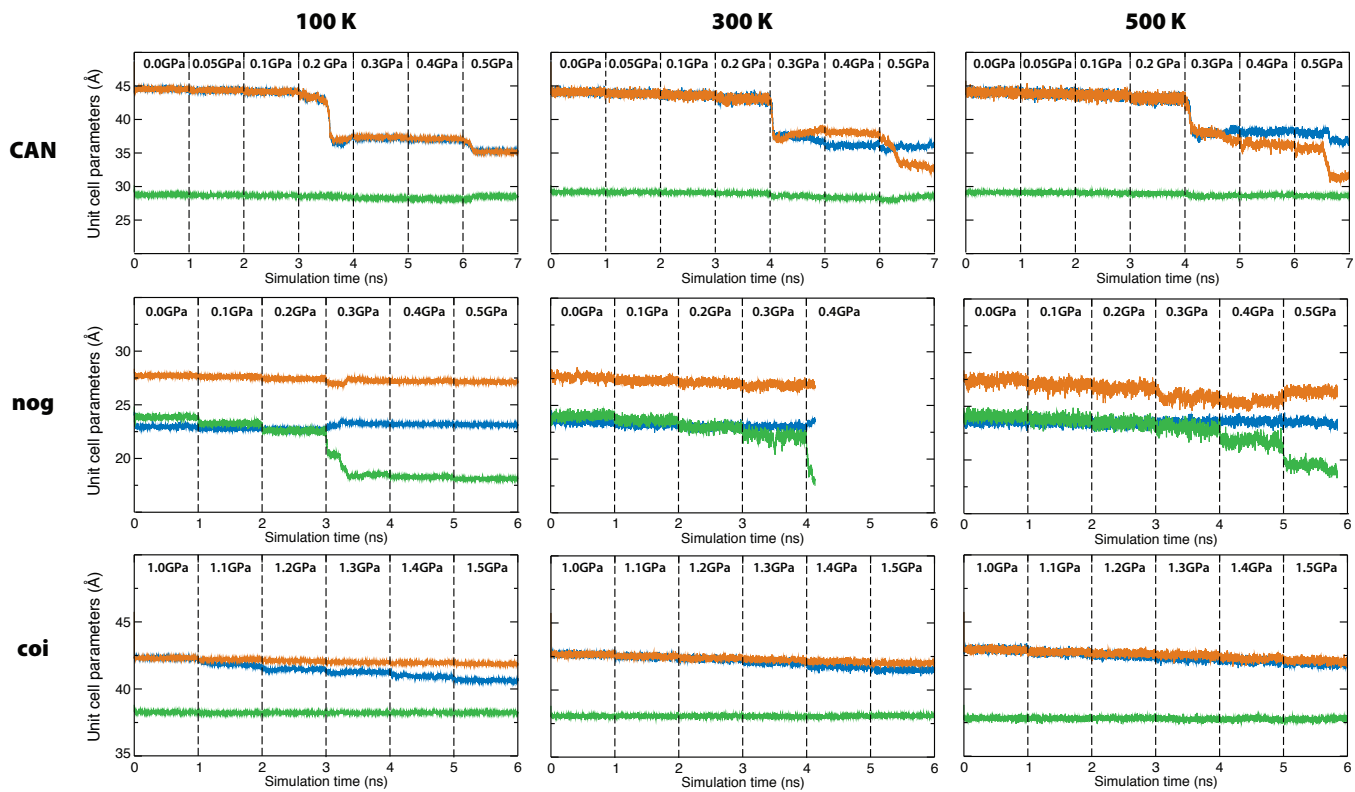

Figure 3. Evolution of the unit cell parameters of ZIFs with **CAN**, **nog** and **coi** topology as a function of increasing pressure (0.1 GPa increment every 1 ns, with an additional point at 0.05 GPa for **CAN**), for temperatures of 100, 300 and 500 K. The curve colors represent unit cell parametrs *a* (blue), *b* (orange), and *c* (green).

| Limit of stability | Materials |
| --- | --- |
| 1.0 GPa | **coi** |
| 0.4 GPa | **SOD**, **nog** |
| 0.3 GPa | **CAN** |
| 0.2 GPa | **MER** |
| 0.05 GPa | **FAU**, **cag**, **LTL** |
| < 0.05 GPa | **AFI**, **DFT** |

Table II. Limits of stability of ZIFs under pressure.

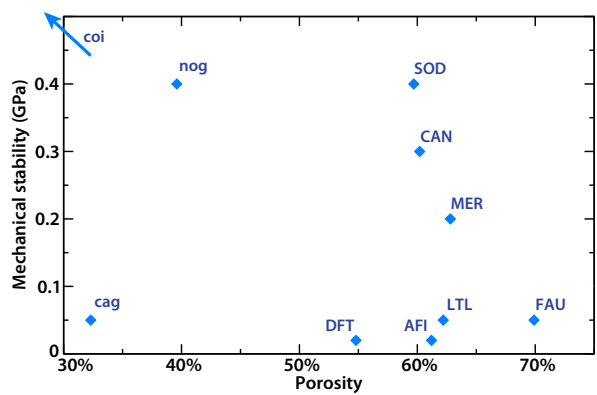

Figure 4. Plot of the mechanical stability (limit of stability, in GPa) against framework porosity (percentage of volume accessible to a spherical probe of radius 1.2 Å).

Moreover, we recently showed that the mechanism behind the instability of ZIF-8 and ZIF-4 under compression is a shear mode softening. That is, isotropic compression induces a lowering of their shear moduli, up to the point where the Born stability conditions no longer hold and the crystal becomes mechanically unstable.[20] In the case of non-cubic crystals, the analysis is made a little bit more difficult because there is not a unique shear modulus, so that one has to look at the evolution of eigenvalues of second-order elastic tensor[27] as a function of pressure. This introduces somewhat larger uncertainties, but qualitative trend is clear nonetheless: all ZIF frameworks studied herein exhibited pressure-induced softening before the point of instability. This mechanism is thus quite generic in the ZIF family of materials, and we suggest it originates from the Zn–*im*–Zn coordination mode itself.

Finally, we looked at the influence of temperature on the stability of three frameworks under pressure. We chose the **CAN**, **nog** and **coi** frameworks, among the most stable, and performed addition compressions experiments, *in silico*, at temperatures of 100, 300 and 500 K. The results are depicted in Figure 3. They show that the influence of temperature in that range is small, but surprisingly higher temperatures appear to give rise to larger stability ranges in pressure. This is in contrast with some of the other frameworks, such as **ATN**, **ATO** and **FER**, where high temperature lead to structural transitions. It might suggest a nonmonotonic or reentrant shape of the temperature–pressure phase diagram for some ZIF structures, and we will perform further work to shed light onto this unexpected effect.

### C. Thermal expansion of ZIFs

In addition to the questions of thermal and mechanical stability discussed above, we also analyze the structural changes of the ten stable ZIF framework as a function of temperature. From MD simulations at various temperatures in the 200–400 K range, we calculate the coefficient of volumetric thermal expansion of each framework,

$$\alpha_V = \frac{1}{V}\left(\frac{\partial V}{\partial T}\right)_\sigma \quad (1)$$

Materials with unusual thermal expansion properties, i.e. very large positive or negative thermal expansion coefficients, or very anisotropic thermal expansion, are highly sought after. Such phenomena can be leveraged for a variety of devices in electronics and optics, sensors and actuators, and in the design of composite materials with zero thermal expansion, for example, dental filling materials.[47]

Figure 5 shows the thermal expansion coefficients calculated in this work, compared to some reference materials from the metal–organic frameworks family: MOF-5[48], MOF-C22[49] and [Ag(en)]NO$_3$[50], all of which have been advertised for their ''exceptional'', ''large'', or ''giant'' (respectively) thermal expansion coefficients (some negative, some positive). We first see that ZIF frameworks show a wide variety of thermal behavior, with some exhibiting positive thermal expansion (PTE) and some negative thermal expansion (NTE). Moreover, the values of the expansion coefficients compare favorably with other metal-organic frameworks, and to dense inorganic materials (typically positive thermal expansion in the range of 10–40 MK$^{-1}$). In particular, we predict two materials with large negative volume thermal expansion (larger than the benchmark MOF-5,[48] and also than molecular framework Zn(CN)$_2$[51], $\alpha_V = -51$ MK$^{-1}$): **LTL** and **MER**, with $\alpha_V = 92$ and 76 MK$^{-1}$ respectively. This prediction is particularly interesting because the **MER** framework has been experimentally synthesized in the Zn(*im*)$_2$ composition, under the name ZIF-10,[1] and its thermal expansion could thus be measured experimentally to test our prediction.

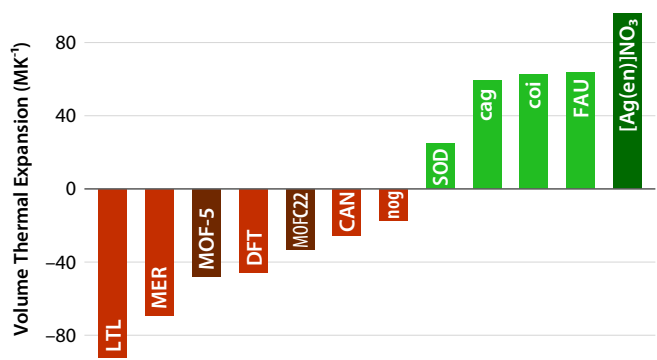

Figure 5. Volume thermal expansion coefficients of the ZIF frameworks studied, compared to values reported in the literature for some other metal–organic frameworks, in darker shade: MOF-5[48], MOF-C22[49] and [Ag(en)]NO$_3$[50].

Finally, it is interesting to note that while the thermal expansion coefficient reported in Figure 5 were calculated in the 200–400 K range, some of the ZIFs with negative thermal expansion (including **nog**, la **DFT** and **LTL**) around room temperature show positive expansion at low temperature, with a transition from PTE to NTE between 100 and 200 K. The same behavior has been previously observed in the Zn(CN)$_2$ molecular framework,[51] with has a four-connected net geometry like ZIFs.

As a conclusion, while ''colossal'' linear thermal expansion (absolute values of the order of $100 \times 10^{-6}$ K$^{-1}$) can be achieved along a single axis of an anisotropic crystal,[52] relying on specific patterns in the mechanical building units of framework materials,[53] we predict that ZIFs have remarkable topology-dependent volume thermal expansion properties, with some achieving large negative thermal expansion compared to known MOFs. Further work on this topic will have to address the links between framework topology and the sign (and extent) of the thermal expansion, as well as microscopic insight into the mechanisms from which these remarkable properties arise.

### IV. CONCLUSION

We performed a systematic study on the thermal and mechanical stability of 18 Zeolitic Imidazolate Frameworks with identical chemical composition and varying framework topology, including known experimental structures and hypothetical structures proposed based on their low enthalpy of formation. We show that many of the hypothetical frameworks proposed in the earlier literature are not mechanical stable at reasonable temperatures (from 77 K to room temperature), in the absence of solvent or guest molecules. Thermal and mechanical stability in working conditions are thus key criteria for the computational prediction of new feasible ZIF structures, and metal–organic frameworks in general, even though they have been little studied so far.

In addition to this study of feasiblity of hypothetical frameworks, we studied the behavior of stable ZIF frameworks upon variations in temperature and pressure. We show that mechanical instability due to pressure-induced elastic softening, which had been demonstrated earlier on ZIF-8 and ZIF-4, is actually a generic feature of the ZIF family. The limits of stability under compression of ZIFs is found to be low in general, with structural transitions occurring at pressures in the 0–400 MPa range for most porous ZIFs. Finally, the analysis of thermal expansion of ZIFs demonstrates a wide variety of behavior as a function of framework topology. Two materials (one experimentally known and one hypothetical) are predicted to show strongly negative volume thermal expansion.

The methodology described here for the assessment of thermal and mechanical stability of hypothetical structures is quite generic, and can readily be used on ZIFs of different chemical composition or other families of metal–organic frameworks (or molecular frameworks) displaying polymorphism. Based on molecular dynamics (MD) simulations at varying temperature and pressure, we have used here classical force field-based MD, since a force field was available in the literature for ZIFs. It could also be done with first principles MD simulations (also known as *ab initio* MD), though at significantly higher computational expense.

### ACKNOWLEDGMENTS

This work was supported by the Agence Nationale de la Recherche under project ''SOFT-CRYSTAB'' (ANR-2010-BLAN-0822), and performed using HPC resources from GENCI-IDRIS (projects 096114 and 100254).